\documentclass[reprint,letterpaper,amssymb,superscriptaddress,nobibnotes,aps, pra]{revtex4-1}
\usepackage{graphicx}
\usepackage{bm}

\begin{document}

\title{Geometrically asymmetric optical cavity for strong atom-photon coupling}

\author{Akio Kawasaki}
\email{akiok@stanford.edu}
\thanks{Present address: W. W. Hansen Experimental Physics Laboratory and Department of Physics, Stanford University, Stanford, California 94305, USA}
\thanks{These authors contributed equally to this work.}
\affiliation{Department of Physics, MIT-Harvard Center for Ultracold Atoms and Research Laboratory of Electronics, Massachusetts Institute of Technology, Cambridge, Massachusetts 02139, USA}

\author{Boris Braverman}
\email{bbraverm@uottawa.ca}
\thanks{Present address: Department of Physics and Max Planck Centre for Extreme and Quantum Photonics, University of Ottawa, 25 Templeton Street, Ottawa, Ontario K1N 6N5, Canada}
\thanks{These authors contributed equally to this work.}
\affiliation{Department of Physics, MIT-Harvard Center for Ultracold Atoms and Research Laboratory of Electronics, Massachusetts Institute of Technology, Cambridge, Massachusetts 02139, USA}

\author{Edwin Pedrozo-Pe\~nafiel}%
\affiliation{Department of Physics, MIT-Harvard Center for Ultracold Atoms and Research Laboratory of Electronics, Massachusetts Institute of Technology, Cambridge, Massachusetts 02139, USA}

\author{Chi Shu}%
\affiliation{Department of Physics, MIT-Harvard Center for Ultracold Atoms and Research Laboratory of Electronics, Massachusetts Institute of Technology, Cambridge, Massachusetts 02139, USA}
\affiliation{Department of Physics, Harvard University, Cambridge, Massachusetts 02138, USA}

\author{Simone Colombo}%
\affiliation{Department of Physics, MIT-Harvard Center for Ultracold Atoms and Research Laboratory of Electronics, Massachusetts Institute of Technology, Cambridge, Massachusetts 02139, USA}

\author{Zeyang Li}%
\affiliation{Department of Physics, MIT-Harvard Center for Ultracold Atoms and Research Laboratory of Electronics, Massachusetts Institute of Technology, Cambridge, Massachusetts 02139, USA}

\author{\"Ozge \"Ozel}%
\affiliation{Department of Physics, MIT-Harvard Center for Ultracold Atoms and Research Laboratory of Electronics, Massachusetts Institute of Technology, Cambridge, Massachusetts 02139, USA}

\author{Wenlan Chen}%
\affiliation{Department of Physics, MIT-Harvard Center for Ultracold Atoms and Research Laboratory of Electronics, Massachusetts Institute of Technology, Cambridge, Massachusetts 02139, USA}

\author{Leonardo Salvi}%
\affiliation{Department of Physics, MIT-Harvard Center for Ultracold Atoms and Research Laboratory of Electronics, Massachusetts Institute of Technology, Cambridge, Massachusetts 02139, USA}
\affiliation{Dipartimento di Fisica e Astronomia and LENS - Universit\`a di Firenze,
INFN - Sezione di Firenze, Via Sansone 1, 50019 Sesto Fiorentino, Italy}

\author{Andr\'e Heinz}%
\thanks{Present address: Max-Planck-Institut f\"ur Quantenoptik, Hans-Kopfermann-Stra\ss
e 1, 85748 Garching, Germany}
\affiliation{Department of Physics, MIT-Harvard Center for Ultracold Atoms and Research Laboratory of Electronics, Massachusetts Institute of Technology, Cambridge, Massachusetts 02139, USA}
\affiliation{Friedrich-Alexander-Universit\"at Erlangen-N\"urnberg, Schlossplatz 4, 91054 Erlangen, Germany}

\author{David Levonian}%
\thanks{Present address: Department of Physics, Harvard University, Cambridge, Massachusetts 02138, USA}
\affiliation{Department of Physics, MIT-Harvard Center for Ultracold Atoms and Research Laboratory of Electronics, Massachusetts Institute of Technology, Cambridge, Massachusetts 02139, USA}

\author{Daisuke Akamatsu}%
\affiliation{Department of Physics, MIT-Harvard Center for Ultracold Atoms and Research Laboratory of Electronics, Massachusetts Institute of Technology, Cambridge, Massachusetts 02139, USA}
\affiliation{National Metrology Institute of Japan (NMIJ), National Institute of Advanced Industrial Science and Technology (AIST), 1-1-1 Umezono, Tsukuba, Ibaraki 305-8563, Japan}

\author{Yanhong Xiao}%
\affiliation{Department of Physics, MIT-Harvard Center for Ultracold Atoms and Research Laboratory of Electronics, Massachusetts Institute of Technology, Cambridge, Massachusetts 02139, USA}
\affiliation{Department of Physics, State Key Laboratory of Surface Physics and Key Laboratory of Micro and
Nano Photonic Structures (Ministry of Education), Fudan University, Shanghai 200433, China}

\author{Vladan Vuleti${\rm{\acute{c}}}$ }
\email{vuletic@mit.edu}
\affiliation{Department of Physics, MIT-Harvard Center for Ultracold Atoms and Research Laboratory of Electronics, Massachusetts Institute of Technology, Cambridge, Massachusetts 02139, USA}
%\texttt{\jobname.tex}
%\date{\today}

\begin{abstract}
Optical cavities are widely used to enhance the interaction between atoms and light. Typical designs using a geometrically symmetric structure in the near-concentric regime face a tradeoff between mechanical stability and high single-atom cooperativity. To overcome this limitation, we design and implement a geometrically asymmetric standing-wave cavity. This structure, with mirrors of very different radii of curvature, allows strong atom-light coupling while exhibiting good stability against misalignment. We observe effective cooperativities ranging from $\eta_{\rm eff}=10$ to $\eta_{\rm eff}=0.2$ by shifting the location of the atoms in the cavity mode. By loading $^{171}$Yb atoms directly from a mirror magneto-optical trap into a one-dimensional optical lattice along the cavity mode, we produce atomic ensembles with collective cooperativities up to $N\eta=2\times 10^4$. This system opens a way to preparing spin squeezing for an optical lattice clock and to accessing a range of nonclassical collective states. 
\end{abstract}

%\pacs{XX, XX, XX}
\maketitle

\section{Introduction}
The interaction between atoms and electromagnetic fields has been studied for more than a century, and has provided many important insights. For an atom at rest, the spectral profile of a single transition is a Lorentzian function. When the atom is so strongly coupled to an electromagnetic mode that its absorption and dispersion appreciably change the mode characteristics, two coupled normal modes with a mixed atom-field character emerge (vacuum Rabi splitting). The strong coupling of an atom to an optical-resonator mode opened the field of cavity quantum electrodynamics (QED) in the optical domain, both for individual atoms \cite{PhysRevLett.68.1132,PhysRevLett.82.3791,PhysRevA.96.031802,RevModPhys.87.1379} and for atomic ensembles \cite{JPhysB.38.S551,Science.333.1266,PhysRevLett.106.133601,Science.344.180,PhysRevLett.99.213601,PhysRevLett.111.100505}. Notable results include the observation of single-atom vacuum Rabi splitting \cite{PhysRevLett.68.1132} and the associated optical nonlinearity \cite{Nature.436.87}, a single-photon transistor \cite{Science.341.768,Nature.536.193,Science.361.57}, a photon-atom quantum gate \cite{Nature.508.237}, polarization-dependent directional spontaneous photon emission \cite{NatCommun.5.5713}, light-induced spin squeezing \cite{Nature.529.505,PhysRevLett.116.093602,PhysRevLett.104.073602,PhysRevLett.113.263603}, preparation of entangled many-atom spin states \cite{Nature.519.439,Science.344.180}, and photon-induced entanglement between distant particles \cite{NatPhoton.8.356}. 

\begin{figure*}[!tb]
	\begin{center}
\includegraphics[width=2.\columnwidth]{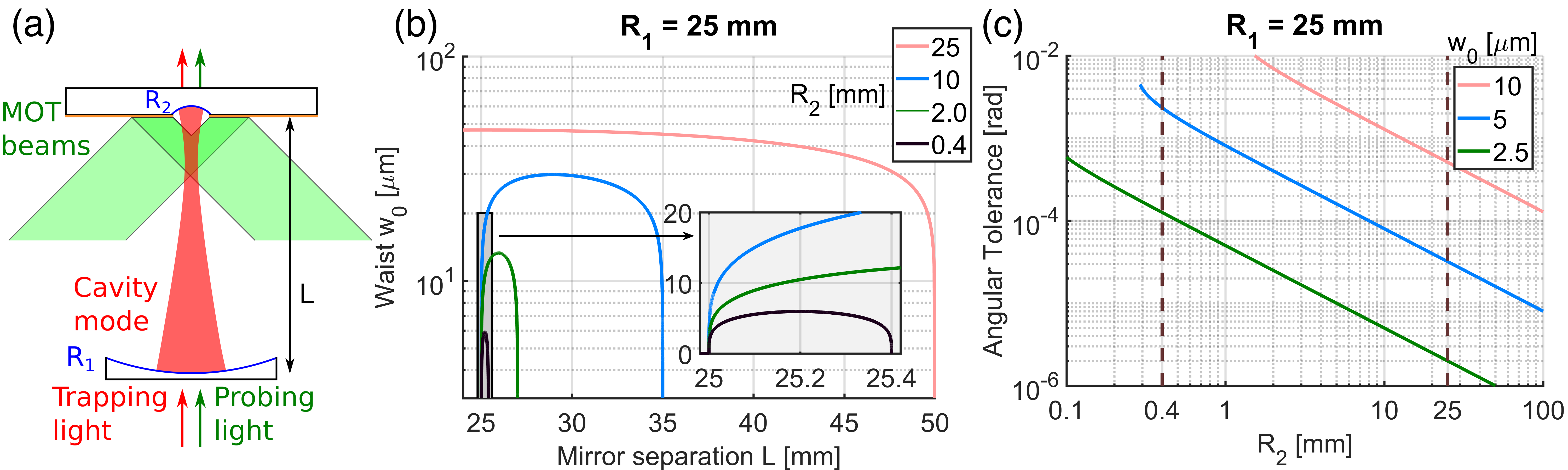}
 \caption{(a) Experimental setup: a cavity mode (dark red) is formed between a mirror of large ROC $R_1$ (bottom) and a micromirror with an ROC of $R_2$ (top), separated by a distance $L$. A mirror MOT (light green) is formed using the flat part of the mirror substrate on which the micromirror is fabricated. Trapping and probing beams are sent through the bottom mirror. 
(b) Waist size $w_0$ of the cavity mode at 556 nm for different values of $R_2$ and mirror separation $L$ when $R_1=25$ mm. The cavity with small $R_2$ permits stable geometries with  small waists.
(c) The angular tolerance $\theta_{\rm T}=D/L$ of the tilt of the optical axis for different $R_2$ and $w_0=10$ $\mu$m, 5 $\mu$m, and 2.5 $\mu$m (top to bottom), fixing $R_1=25$ mm. The asymmetric cavity with $R_1 \gg R_2$ is far more stable with respect to misalignment than the near-concentric symmetric cavity with $R_1 = R_2\approx L/2$ for a given $w_0$.
} 
 \label{IntroFig}
 \end{center}
\end{figure*}

The most common structure used in cavity QED experiments is a Fabry-Perot (FP) cavity consisting of two spherical mirrors with equal radii of curvature (ROCs) \cite{JPhysB.38.S551,Science.333.1266,PhysRevLett.106.133601,Science.344.180,PhysRevLett.99.213601}. For confocal and shorter cavities, this configuration exhibits good mechanical stability of the optical mode. However, when it comes to increasing the single-atom cooperativity $\eta$, the structure has certain constraints: to achieve a small mode waist with commercially available super-polished mirrors of centimeter-scale ROCs, the two mirrors need to be very far from each other (near-concentric cavity) \cite{ApplPhysB.107.1145}, or very close to each other (near-planar cavity) \cite{JPhysB.38.S551}. The near-concentric cavity is very sensitive to alignment errors, while the near-planar cavity offers little optical access. To overcome these difficulties, we instead implemented a geometrically asymmetric cavity, which offers good optical access and a very small mode waist at reasonable mechanical stability. This paper describes the concept and experimental realization of such an asymmetric cavity with high $\eta$. We observe single-atom cooperativity up to $\eta=10$, and collective cooperativity up to $N\eta=2\times 10^4$ with trapped-atom lifetime exceeding several seconds. 

\section{Concept of asymmetric cavity}
Cavity QED is a gateway for manipulating single atoms and atomic ensembles using light \cite{PhysScripta.T76.127,AdvAtMolOptPhys.60.201}. The all-important parameter is the single-atom cooperativity at an antinode $\eta_{\rm max}$, given by 
\begin{equation}\label{EqCooperativity}
\eta_{\rm max}=\frac{4g^2}{\kappa\Gamma}=\frac{24 {\cal F}}{\pi k^2 w^2}
\end{equation}
for a standing wave cavity \cite{AdvAtMolOptPhys.60.201}. This parameter is a dimensionless constant in cavity QED that describes the strength of atom-light interaction, where $2g$ is the coupling constant (single-photon Rabi frequency) between an atom and a photon, $\kappa$ is the decay rate of a photon in the cavity, $\Gamma$ is the decay rate of the atomic excited state, ${\cal F}$ is the finesse of the cavity, $k=2\pi/\lambda$ is the wavenumber, and $w$ is the $1/e^2$ intensity radius of the cavity mode. An important realization in cavity QED is that the ratio of the coupling constant squared and the product of decay rates is purely geometric. Therefore, designing a cavity with $\eta \gg 1$, useful for obtaining highly entangled states using light \cite{PhysRevLett.115.250502}, is reduced to designing a cavity with small beam size $w$ and high finesse ${\cal F}$. 

The geometrical relation between the ROCs and positions of two mirrors, and the resulting shape of the cavity mode are well known (e.g. \cite{Siegman}). If one uses more than two mirrors, a waist size smaller than that with a conventional two-mirror cavity can be realized \cite{JPhysB.51.195002}, but here we concentrate on a cavity with two mirrors, because it benefits from a simpler mechanical structure and lower optical loss. In the general case, the waist size for a two-mirror cavity is given by \cite{Siegman}
\begin{equation}\label{EqWaist}
w_0^2=\frac{L\lambda}{\pi}\sqrt{\frac{g_1g_2\left( 1-g_1g_2 \right)}{\left( g_1+g_2-2g_1g_2\right)^2}},
\end{equation}
where $g_{1,2}=1-L/R_{1,2}$, $R_{1,2}$ denote the ROC of the two mirrors, and $L$ is the distance between the two mirrors. 

In the case of a symmetric cavity ($R_1=R_2$ and thus $g_1=g_2$), this expression simplifies to $w_0^2=(L\lambda/2\pi)\sqrt{(1+g_1)/(1-g_1)}$, leading to two possible cavity configurations with small $w_0$: (i) when the two mirrors are very close to each other, $L\approx 0$, and (ii) when the two mirrors are in a near-concentric configuration, $L\approx 2R_1$. 

The first configuration has good mechanical stability due to a large optical axis length, given by the distance $D=R_1+R_2-L \approx 2R_1$ between the centers of curvature of the two mirrors. This is a good configuration for having very high cooperativity, and has been used in many experiments, particularly with single atoms \cite{PhysRevLett.68.1132,PhysRevLett.82.3791,Nature.436.87,JPhysB.38.S551}, though the optical access for loading and manipulating atoms is very limited. With additional technical effort, such as a movable magnetic trap \cite{PhysRevLett.99.213601,Nature.450.272,PhysRevLett.98.233601}, it is possible to load large atomic ensembles even into very short cavities. 

The near-concentric configuration, on the other hand, offers excellent optical access for loading atoms directly into the cavity mode from a magneto-optical trap (MOT) or any other type of trap. However, in this case, the length of the optical axis is short: $D=2\pi^2 w_0^4/(R_1 \lambda^2)$. For example, to obtain $w_0=5$ $\mu$m with $R_1=25$ mm, the cavity has $D=1.6$ $\mu$m for 556 nm light. This causes difficulties in obtaining and maintaining alignment of the cavity, as well as poor mechanical stability. In this case, higher-order transverse modes are close to the fundamental mode in frequency, which can be problematic for experiments aiming to couple atoms to a single cavity mode. Nevertheless, this type of cavity is used for ions to keep the mirror surfaces far away from the trapped particles \cite{PhysRevLett.111.100505}. Some cavities even utilize mirrors with aspheric structure to attain the large numerical aperture required for focusing the beam tightly \cite{NewJPhys.16.103002,1806.03038}. 

Next, we consider an asymmetric cavity with $R_1 \gg R_2$ [see Fig. \ref{IntroFig}(a)]. In this case, there are two separate stability regions, one with $0<L<R_2$ and the other with $R_1<L<R_1+R_2$. Figure \ref{IntroFig}(b) shows the waist size in the long stability region with $L>R_1$. As $R_2$ shrinks, so does the maximum mode waist $w_0$. When $R_1$ and $R_2$ are fixed, larger $w_0$ gives a larger angular tolerance $\theta_T=D/L$, which is the sensitivity of the optical axis alignment to any tilt in the cavity mounting hardware. When a target $w_0$ is set and $R_2$ is varied, smaller $R_2$ gives larger angular tolerance $\theta_T$, as shown in Fig. \ref{IntroFig}(c). This motivates the construction of an asymmetric cavity consisting of a standard super-polished mirror of $R_1=25$ mm and a micromirror of $R_2\sim400$ $\mu$m, which is manufactured by ablation with a CO$_2$ laser pulse \cite{NJP12.065038}, to simultaneously achieve high cooperativity, large distance between the two mirrors, and large angular tolerance $\theta_T$. Compared to a symmetric cavity with $R_1=R_2=25$ mm, this setup is 60 times more stable with respect to angular misalignment [see Fig. \ref{IntroFig}(c)].

\section{Cavity properties}\label{CavProperty}
\begin{figure}[!t]
	\begin{center}
 \includegraphics[width=0.8\columnwidth]{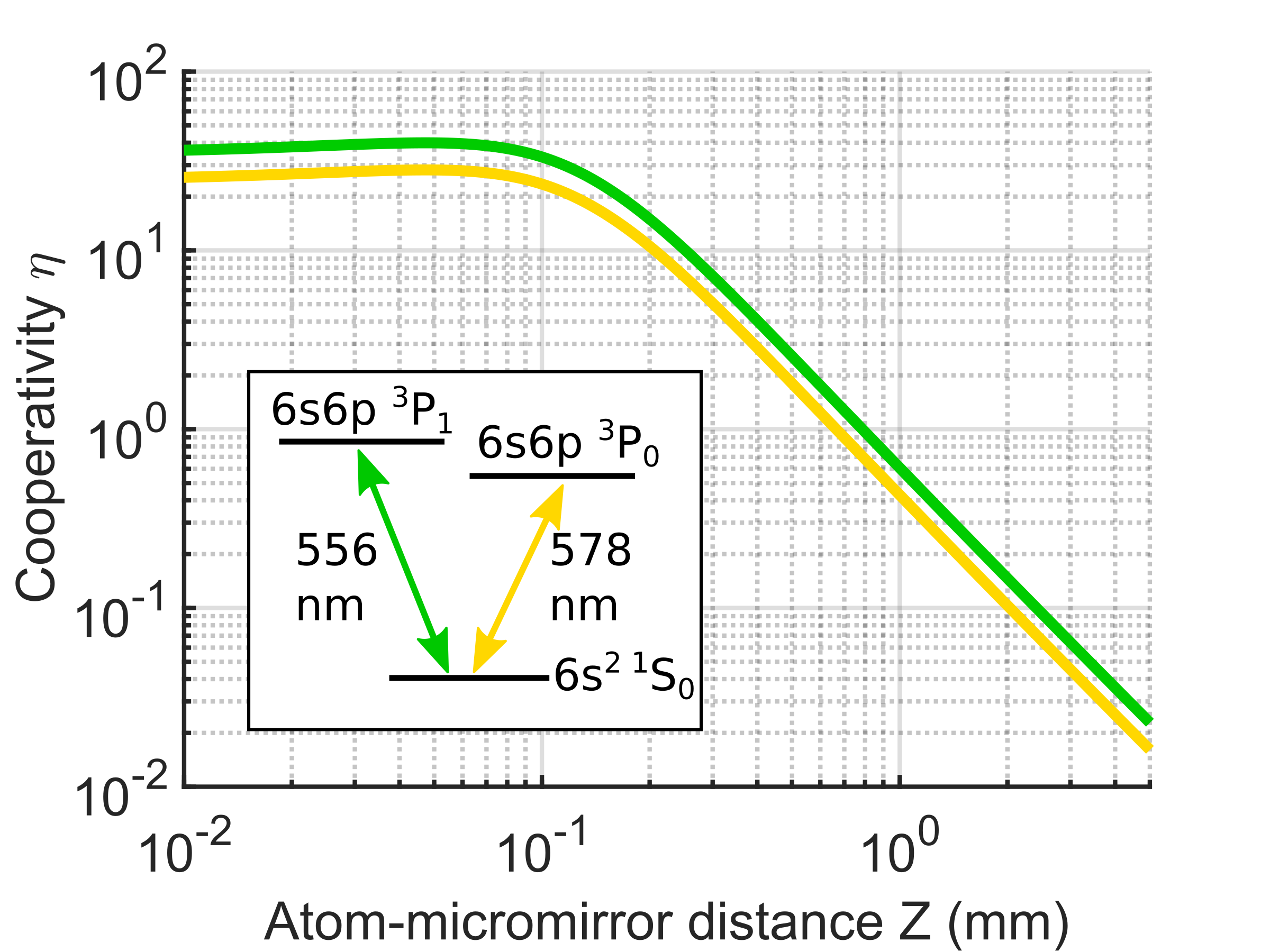}
 \caption{Single-atom cooperativity at the cavity mode antinodes, calculated from the geometry and the finesse ${\cal F}$ for the atoms trapped in the cavity at different locations along the cavity axis. The upper green curve corresponds to the $\mathrm{6s^2 \hspace{0.1 em} {}^{1} \hspace{-0.05 em} S_0 \rightarrow 6s6p \hspace{0.1 em}{} ^{3} \hspace{-0.05 em} P_1}$ transition at 556 nm for ${\cal F}=1.4 \times 10^4$, and the lower yellow curve corresponds to the $\mathrm{6s^2 \hspace{0.1 em} {}^{1} \hspace{-0.05 em} S_0 \rightarrow 6s6p \hspace{0.1 em}{} ^{3} \hspace{-0.05 em} P_0}$ clock transition at 578 nm for ${\cal F}=9.5 \times 10^3$.}
 \label{etavsatompos}
 \end{center}
\end{figure}

We built an asymmetric cavity with a slightly elliptical micromirror ($R_{\rm 2x}=303$ $\mu$m, $R_{\rm 2y}=391$ $\mu$m \cite{BBthesis}) on a flat substrate and a standard super-polished mirror ($R_1=25$ mm, see Appendixes for the mechanical details and the procedure of construction). The mirrors have high reflectivity coatings for 556 nm and 759 nm light at normal incidence. The mirrors also reflect 99 \% of the 399 nm and 556 nm light at 45$^{\circ}$ angle of incidence to enable the operation of a mirror MOT with ytterbium, as shown in Fig. \ref{IntroFig}(a). Prior to fixing the mirror distance, the finesse ${\cal F}$ is measured for different separations between the two mirrors.  A constant ${\cal F}$ is observed in the region of $25.00<L<25.12$ mm, and it decreases at larger $L$, which may be caused by extra loss due to the large mode size on the nonspherical micromirror \cite{NewJPhys.17.053051}. The inter-mirror distance is fixed at $L=25.0467(10)$ mm, which is calibrated by the disappearance of the cavity mode when $L<R_1$ and a known shift by a micrometer stage. Note that this distance is different from $L=25.10807(17)$ mm derived from the measured free spectral range (FSR) of 5970.04(4) MHz, which potentially implies the breakdown of the simple relation between the FSR and cavity length at small waist size, where the paraxial approximation no longer holds (see Appendix \ref{AssymCavFSR} for more discussion). The expected cooperativity $\eta$ for different atom position $Z$, defined as the distance of the atoms from the micromirror, is calculated based on the mode geometry and ${\cal F}$. The single-atom cooperativity $\eta$ and other QED parameters are summarized in Table \ref{cavityqedparameter} and Fig. \ref{etavsatompos}. In addition to the $\mathrm{6s^2 \hspace{0.1 em} {}^{1} \hspace{-0.05 em} S_0 \rightarrow 6s6p \hspace{0.1 em}{} ^{3} \hspace{-0.05 em} P_1}$ transition at 556 nm, the cavity also has a high finesse for the $\mathrm{6s^2 \hspace{0.1 em} {}^{1} \hspace{-0.05 em} S_0 \rightarrow 6s6p \hspace{0.1 em}{} ^{3} \hspace{-0.05 em} P_0}$ clock transition at 578 nm. The single-atom cooperativity $\eta$ for 556 nm light can be tuned from the maximum of 40 to less than 0.1 by changing the position of the atoms by a few millimeters, as shown in Fig. \ref{etavsatompos}. 

\begin{table}[!t]
\caption{Cavity QED parameters of the constructed cavity for 556, 578, and 759 nm light, corresponding to the $\mathrm{6s^2 \hspace{0.1 em} {}^{1} \hspace{-0.05 em} S_0 \rightarrow 6s6p \hspace{0.1 em}{} ^{3} \hspace{-0.05 em} P_1}$ transition, the $\mathrm{6s^2 \hspace{0.1 em} {}^{1} \hspace{-0.05 em} S_0 \rightarrow 6s6p \hspace{0.1 em}{} ^{3} \hspace{-0.05 em} P_0}$ clock transition, and the magic wavelength for the clock transition, respectively. $R_{\rm 25mm}$ and $R_{\rm micro}$ are reflectivities for the 25 mm ROC mirror and the micromirror, respectively, and ${\cal F}$ is the corresponding finesse. }
\begin{center}
\begin{tabular}{cccc}
wavelength $\lambda$ & 556 nm & 578 nm & 759 nm \\
\hline
$1-R_{\rm 25mm}$ & 60(2) ppm & 80(5) ppm & 1000(50) ppm \\
$1-R_{\rm micro}$ & 390(10) ppm & 580(20) ppm & 1000(50) ppm \\
${\cal F}/10^3$ & $14.0(1)$ & $9.5(1)$ & $3.14(7)$\\
$\Gamma/(2\pi)$ & 184(1) kHz & 7.0(2) mHz & - \\
$\kappa/(2\pi)$ & 426(2) kHz & 628(4) kHz & 1.90(4) MHz\\
$g_{\rm max}/(2\pi)$ & 885(5) kHz & 176(1) Hz & -\\
$\eta_{\rm max}$ & 40.0 & 28.2 & - \\
$w_0$ & 4.60 $\mu$m & 4.70 $\mu$m & 5.38 $\mu$m 
\end{tabular}
\end{center}
\label{cavityqedparameter}
\end{table}%

\section{Atom trapping in the cavity mode}
To measure the single-atom cooperativity $\eta$ with atoms, a mirror MOT \cite{PhysRevLett.83.3398} is operated with $^{171}$Yb [see also Fig. \ref{IntroFig}(a)]. The atoms are first loaded into a two-color MOT \cite{JPhysB.48.155302}. Subsequently, the 399 nm cooling light on the $\mathrm{6s^2 \hspace{0.1 em} {}^{1} \hspace{-0.05 em} S_0 \rightarrow 6s6p \hspace{0.1 em}{} ^{1} \hspace{-0.05 em} P_1}$ transition is turned off, the detuning of the 556 nm MOT light is reduced from $-7$ MHz to $-200$ kHz (the linewidth of the transition is $\Gamma = 2\pi\times 184$ kHz), and a bias magnetic field is added to move the atoms to the desired location along the cavity axis. Typically around $10^4$ $^{171}$Yb atoms are trapped in the MOT by 556 nm light at a temperature of 15 $\mu$K, with a rms cloud radius of 60 $\mu$m along the vertical cavity axis. 

To trap the atoms in the cavity mode, a one-dimensional optical lattice near the magic wavelength of 759 nm for the clock transition is generated inside the cavity. With a typical circulating power of 1.2 W, the trap depth at a distance of $Z=0.42$ mm from the micromirror is $2.5$ MHz, %this corresponds to 50 uK. note that 50 uK corresponds to 500 Er and Er=3.4 kHz
with trapping frequencies 142(3) kHz axially and 1.39(10) kHz radially. To load the atoms into the optical lattice, the detuning of the 556 nm MOT light is increased from $-200$ kHz to $-400$ kHz, and the intensity per beam is lowered to 0.05 mW/cm$^2$ (the saturation intensity of the transition is 0.14 mW/cm$^2$) for 20 ms before the MOT light is extinguished. The lifetime of the atoms in the optical lattice is typically a few seconds, limited by intensity noise in the lattice, and approaching the limit set by background gas collisions. 

\section{Single-atom and collective cooperativity measurement}
A cavity-QED system with atoms in the cavity mode is typically characterized by the single-atom cooperativity $\eta$ and the collective cooperativity $N\eta$, where $N$ is the atom number. The single-atom cooperativity $\eta$ determines the strength of the interaction between atoms and light, while the collective cooperativity $N\eta$ sets some limits for the manipulation of the quantum system, such as the amount of attainable spin squeezing (e.g., \cite{PhysRevA.81.021804,PhysRevA.89.043837}). This is because $N\eta$ determines the ratio of useful collective light scattering by the ensemble into the cavity relative to the scattering of light into free space, which results in decoherence \cite{AdvAtMolOptPhys.60.201}.

\begin{figure}[!t]
	\begin{center}
 \includegraphics[width=0.7\columnwidth]{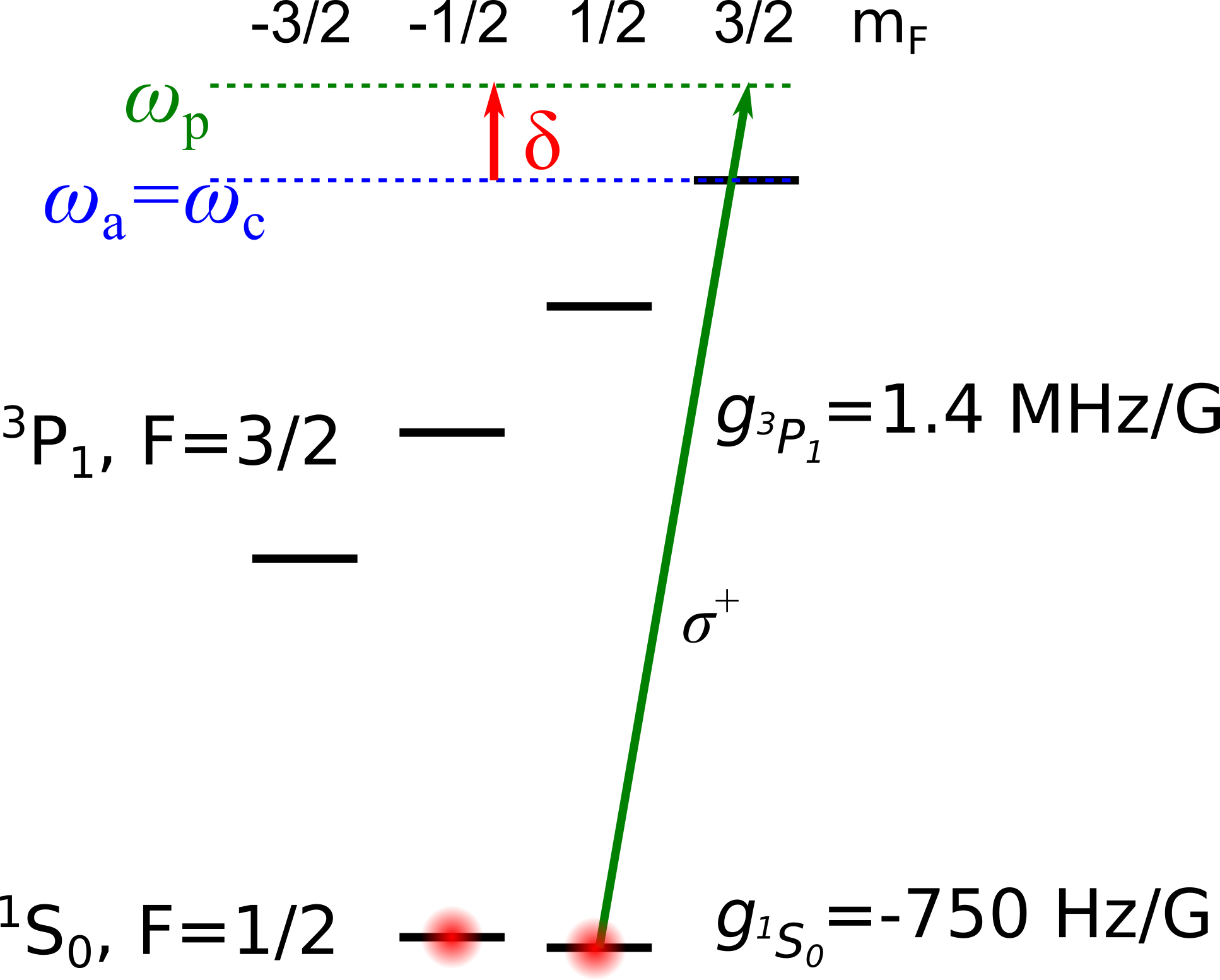}
 \caption{Hyperfine structure of the $\mathrm{6s^2 \hspace{0.1 em} {}^{1} \hspace{-0.05 em} S_0}$ state and the $\mathrm{6s6p \hspace{0.1 em}{} ^{3} \hspace{-0.05 em} P_1}$ state relevant to phase shift measurement: the $F=1/2$ manifold of the $\mathrm{6s6p \hspace{0.1 em}{} ^{3} \hspace{-0.05 em} P_1}$ state is $-6$ GHz detuned from the $F=3/2$ manifold and therefore is not drawn in the figure. $g_{3P1}$ and $g_{1S0}$ are g factors for the $\mathrm{6s6p \hspace{0.1 em}{} ^{3} \hspace{-0.05 em} P_1}$ state and the $\mathrm{6s^2 \hspace{0.1 em} {}^{1} \hspace{-0.05 em} S_0}$ state, respectively. $\omega_{\rm a}$, $\omega_{\rm c}$, and $\omega_{\rm p}$ are the frequencies of the $\mathrm{6s^2 \hspace{0.1 em} {}^{1} \hspace{-0.05 em} S_0 \rightarrow 6s6p \hspace{0.1 em}{} ^{3} \hspace{-0.05 em} P_1}$ atomic transition, the cavity resonance, and the probing laser, respectively.}
 \label{EnergyDiagramForPhase}
 \end{center}
\end{figure} 

\subsection{Single-atom cooperativity} 
\begin{figure}[!t]
	\begin{center}
 \includegraphics[width=0.8\columnwidth]{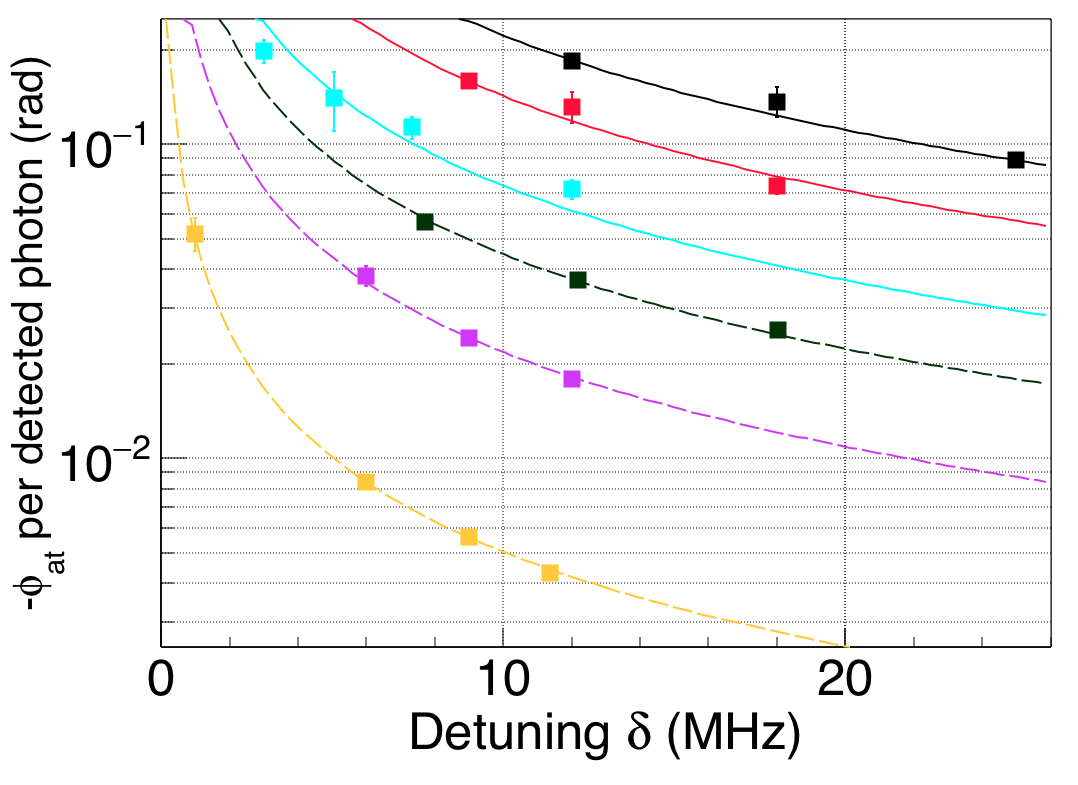}
 \caption{Phase shift measurements at $Z=0.14$, 0.27, 0.44, 0.558, 0.564, and 1.40 mm (from top to bottom): squares are the measured phase shifts at different detunings $\delta$, and curves are the fitted phase shift, including the effect of both $|m_F=\pm1/2\rangle$ states.}
 \label{PhaseMeasurement}
 \end{center}
\end{figure}

The single-atom cooperativity $\eta$ %, measured as an average of $\eta$ for all atoms trapped in the cavity mode, 
can be experimentally determined as the effective single-atom cooperativity $\eta_{\rm eff}$ by measuring the atomic phase shift $\phi_{\rm at}$ induced by off-resonant probing light \cite{AdvAtMolOptPhys.60.201}. The measured value of $\eta_{\rm eff}$ equals $(3/4)\eta_{\rm max}$, assuming a uniform distribution of atoms along the cavity mode \cite{PhysRevA.92.063816}. To perform the measurement, atoms are optically pumped into the $|^1S_0, m_F=+1/2\rangle$ state, with a bias magnetic field $B=13.6$ G parallel to the cavity axis applied to generate an energy difference of $h \times 10.2$ kHz between the $|^1S_0, m_F= \pm 1/2\rangle$ states, where $h$ is the Planck constant (see Fig. \ref{EnergyDiagramForPhase} for the detailed energy level structure of the system). The cavity resonance frequency $\omega_{\rm c}$ is set equal to the atomic resonance frequency $\omega_{\rm a}$ for the $|\mathrm{6s^2 \hspace{0.1 em} {}^{1} \hspace{-0.05 em} S_0,m_F=+1/2 \rangle \rightarrow |6s6p \hspace{0.1 em}{} ^{3} \hspace{-0.05 em} P_1, m_F=+3/2} \rangle$ transition, and the probing light is detuned by $\delta$ from both resonances. After applying a $\pi/2$ pulse to the atoms resonant with the Zeeman splitting of the ground state, a probing laser pulse is sent into the cavity mode, which shifts the phase between the $|m_F=\pm1/2\rangle$ states by an amount 
\begin{equation}\label{EqPhaseShift}
\phi_{\rm at}=-\frac{\eta_{\rm eff}}{2\epsilon} \frac{2\delta/\Gamma}{1+(2\delta/\Gamma)^2}
\end{equation}
per detected photon. The system quantum efficiency $\epsilon$ is defined as $\epsilon=(1-L_{\rm op})\frac{T_2}{T_1+T_2+L_1+L_2}$, where $T_1$ and $T_2$ are the transmission of the input- and output-side mirrors, $L_1$ and $L_2$ are the loss at the input- and output-side mirrors, and $L_{\rm op}$ is the loss between the output-side mirror and the photodetector including the detector's quantum efficiency. \cite{BBthesis,AdvAtMolOptPhys.60.201}. The phase is measured as a population difference between the $|m_F=\pm1/2\rangle$ states after another $\pi/2$ pulse. Figure \ref{PhaseMeasurement} shows the result of the phase measurements, including the small additional phase shift from the $|\mathrm{6s^2 \hspace{0.1 em} {}^{1} \hspace{-0.05 em} S_0,m_F=-1/2 \rangle \rightarrow |6s6p \hspace{0.1 em}{} ^{3} \hspace{-0.05 em} P_1, m_F=+1/2} \rangle$ transition. The measurements at different detunings $\delta$ are fitted reasonably well by Eq. (\ref{EqPhaseShift}) with $\eta_{\rm eff}/\epsilon$ as the only fitting parameter. From these fits, the cooperativity $\eta_{\rm eff}$ at different atom positions is calculated, assuming the overall detection efficiency of an intracavity photon $\epsilon$ is 0.175(30), obtained from independent measurements of the cavity and photodetector properties. Note that the uncertainty of $\epsilon$ propagates into the estimate of $\eta_{\rm eff}$ as a systematic error. 

%original Zs $Z=0.1$, 0.2, 0.32, 0.41, 0.415, and 1.03 mm
%new Zs $Z=0.16$, 0.32, 0.51, 0.656, 0.664, and 1.65 mm
%correct new Zs: Z=0.14, 0.27, 0.44, 0.558, 0.564, 1.40

The measured effective single-atom cooperativity in this system ranges from $\eta_{\rm eff}=10$ to $\eta_{\rm eff}=0.2$ for atom-micromirror distances between $Z=0.136$ mm and $Z=1.40$ mm as shown in Fig. \ref{CooperativityResult}. The value of $Z$ has systematic uncertainty of 7\% due to uncertainty in the magnification of the imaging system. The measured effective cooperativity matches well with the calculated value, as shown in Fig. \ref{CooperativityResult}.
%Compared to the calculated value (Fig. \ref{etavsatompos}), the measured effective cooperativity is lower by a factor of 2, presumably due to the finite atom temperature that distributes atoms to a broader region including where the atom-light coupling is small, reducing the average coupling to the cavity mode. 

\subsection{Collective cooperativity}
To measure the collective cooperativity $N\eta$ after trapping the atoms inside the cavity, we measure the vacuum Rabi splitting of the cavity resonance $\Delta\omega$. $N\eta$ is given by \cite{AdvAtMolOptPhys.60.201}
\begin{equation}
N\eta=\frac{(\Delta\omega)^2}{\kappa\Gamma}
\end{equation}
For the measurement of $\Delta\omega$, the atomic and cavity resonances are set to the same frequency $\omega_{\rm a}=\omega_{\rm c}$, and a probing laser at 556 nm is sent into the system. The vacuum Rabi splitting $\Delta\omega$ is obtained by the phase and the power measurement of the transmitted probing laser whose frequency $\omega_{\rm p}$ is scanned over the resonance peak. The scanning is performed by two sidebands $\omega_{\rm p} = \omega_{\rm a} \pm \omega_{\rm ch}$ to cancel the effect of the fluctuation of the cavity resonance frequency under the condition of $\Delta\omega \gg \kappa,\Gamma$, where the chirping frequency $\omega_{\rm ch}$ increases linearly in time. 

\begin{figure}[!t]
	\begin{center}
 \includegraphics[width=0.8\columnwidth]{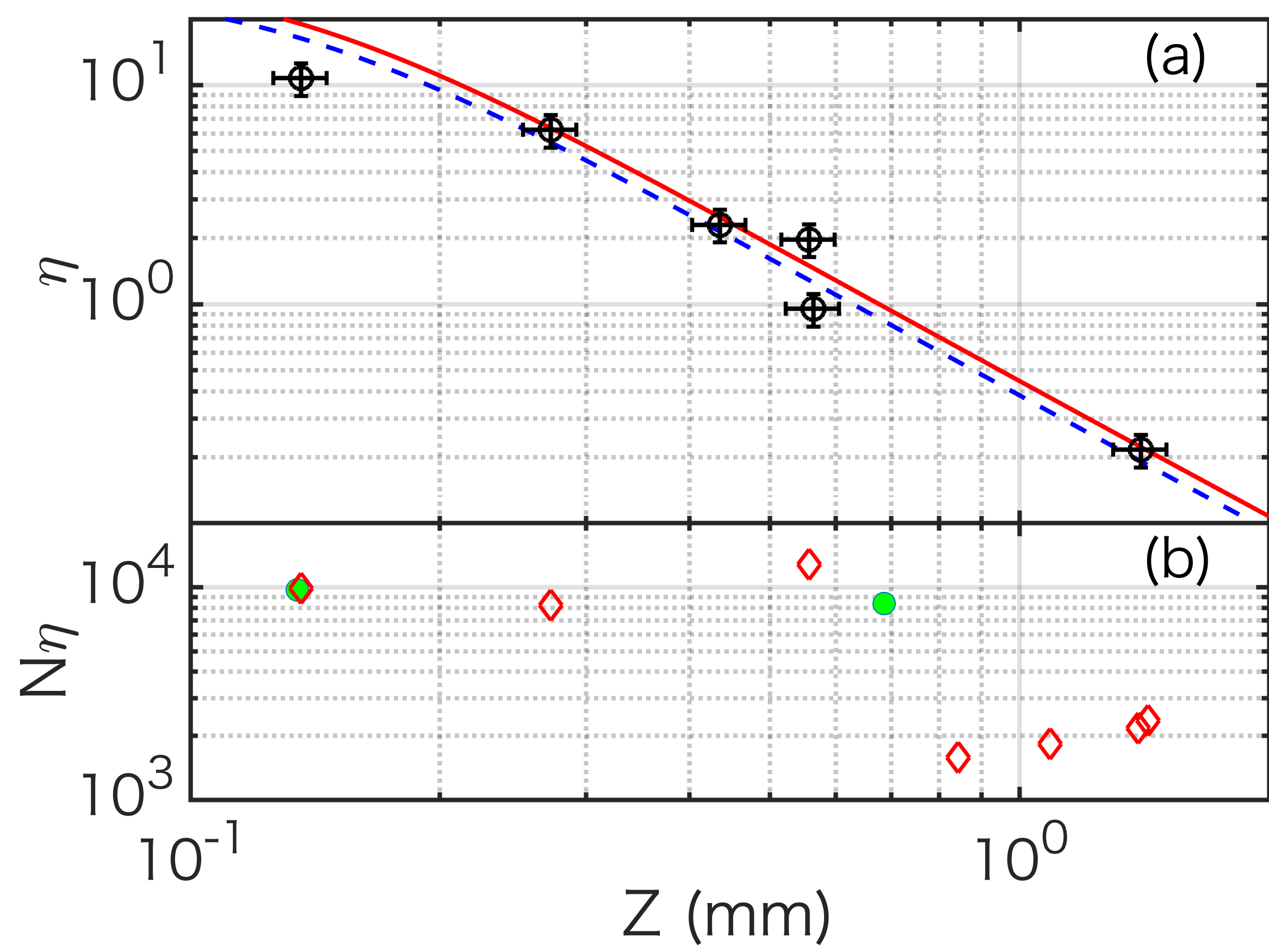}
 \caption{(a) Measured effective single-atom cooperativity $\eta_{\rm eff}$ and (b) collective cooperativity $N\eta$ for different atom distances $Z$ from the micromirror. (a) The black circles are the measured $\eta_{\rm eff}$. The error bars show the systematic error, while the statistical error is negligible. The solid red curve is the $\eta_{\rm eff}$ estimated from the geometry of the cavity shown in Fig. \ref{etavsatompos}, and dashed blue curve is the best fit of the measured $\eta_{\rm eff}$. (b) Collective cooperativity $N\eta$ measured via vacuum Rabi splitting (red diamonds) or cavity frequency shift (green circles).}
 \label{CooperativityResult}
 \end{center}
\end{figure}

 Alternatively, one can also measure $N\eta$ by measuring the dispersive shift of cavity resonance frequency $\delta \omega_c$, according to the following equation:
\begin{equation}\label{EqOffRes}
N\eta=\delta\omega_{\rm c} \frac{4\Delta}{\kappa\Gamma}
\end{equation}
To perform this frequency shift measurement, $\omega_{\rm p}$ is fixed as $\Delta=\omega_{\rm p}-\omega_{\rm a}$ and the relative transmission through the cavity is measured. The values of $N\eta$ derived from both methods agree with each other. Fig. \ref{CooperativityResult}(b) shows that collective cooperativities $N\eta$ up to $10^4$ are observed for a wide range of atom positions $Z$. The observed values of $N\eta$ are sufficiently large to permit significant cavity-feedback or measurement-based spin squeezing \cite{PhysRevA.81.021804,AKThesis,BBthesis} in future experiments. The details of atom trapping to a small optical lattice are discussed elsewhere \cite{TrappingPaper}.

\section{Summary}

We have constructed an asymmetric cavity reaching the single-atom strong-coupling regime, and have measured a cooperativity up to $\eta_{\rm eff}=10$ for $^{171}$Yb atoms on the $\mathrm{6s^2 \hspace{0.1 em} {}^{1} \hspace{-0.05 em} S_0 \rightarrow 6s6p \hspace{0.1 em}{} ^{3} \hspace{-0.05 em} P_1}$ transition. The asymmetric structure with a standard mirror and a micromirror ensures both large single-atom cooperativity and mechanical stability, as well as easy tuning of cooperativity by changing the atom position. Atom trapping is performed by a mirror MOT, and collective cooperativities $N\eta$ in excess of $10^4$ are reached at atom-micromirror distances $Z\leq0.7$ mm in a one-dimensional optical lattice with a lifetime exceeding 1 s. The measured single-atom cooperativity ranges from $\eta=10$ to $\eta=0.2$, in agreement with the value expected from the cavity geometry and finesse. The large collective cooperativity we observe will enable spin squeezing in the $|m_F=\pm1/2\rangle$ ground-state manifold, which can then be mapped onto the atomic clock transition, as well as preparation of non-classical collective states \cite{RevModPhys.90.035005}.

\begin{acknowledgments}
This work is supported by DARPA Grant No. W911NF-11-1-0202, NSF Grants No. PHY-1505862 and No. PHY-1806765, NSF CUA Grant No. PHY-1734011, ONR Grant No. N00014-17-1-2254, and AFOSR MURI Grant No. FA9550-16-1-0323. B.B. acknowledges support from the National Science and Engineering Research Council of Canada. 

A.K. and B.B. contributed equally to this work. 
\end{acknowledgments}

\appendix
\section{How to assemble the asymmetric cavity with micromirror}\label{AppA}
The construction of the asymmetric cavity with a micromirror has to follow a specific procedure \cite{BBthesis}, since the center of the large-ROC mirror has to be precisely aligned into a cone of 100 $\mu$m diameter and 300 $\mu$m height consisting of the micromirror and its center of curvature. 

First, without the micromirror, the large-ROC mirror is aligned to the light that is sent from the micromirror side. The alignment is performed by matching the retroreflected light to the path of the incident laser beam. This aligns the large-ROC mirror to the optical axis of the cavity set by the input beam. Next, the micromirror is inserted. To do this, the flat part of the micromirror substrate is first used to make a cavity with the large-ROC mirror (flat-large ROC cavity), with transmission monitored by a CCD camera. If a cavity is formed, discrete transmission peaks corresponding to different transverse Hermite-Gaussian modes are observed. After the input light is aligned to couple mainly to the fundamental mode, the micromirror substrate is moved farther and farther from the large ROC mirror, until the cavity mode disappears. This ensures that the distance between the two mirrors is exactly the same as the ROC of the large-ROC mirror. 

The third step is to align the transverse position of the micromirror substrate. This is performed simply by translating the substrate until strong scattered light from the micromirror is observed. At this point, the transmission often has two spots, corresponding to the cavity mode in a V-shaped configuration, with two points of reflection on the micromirror substrate. The goal is to merge these two spots into one, and this is the situation where a good cavity mode is formed for the asymmetric cavity.

\section{Details of the mechanical structure of the asymmetric cavity}\label{AssymCavFSR}
\begin{figure}[!t]
	\begin{center}
 \includegraphics[width=1\columnwidth]{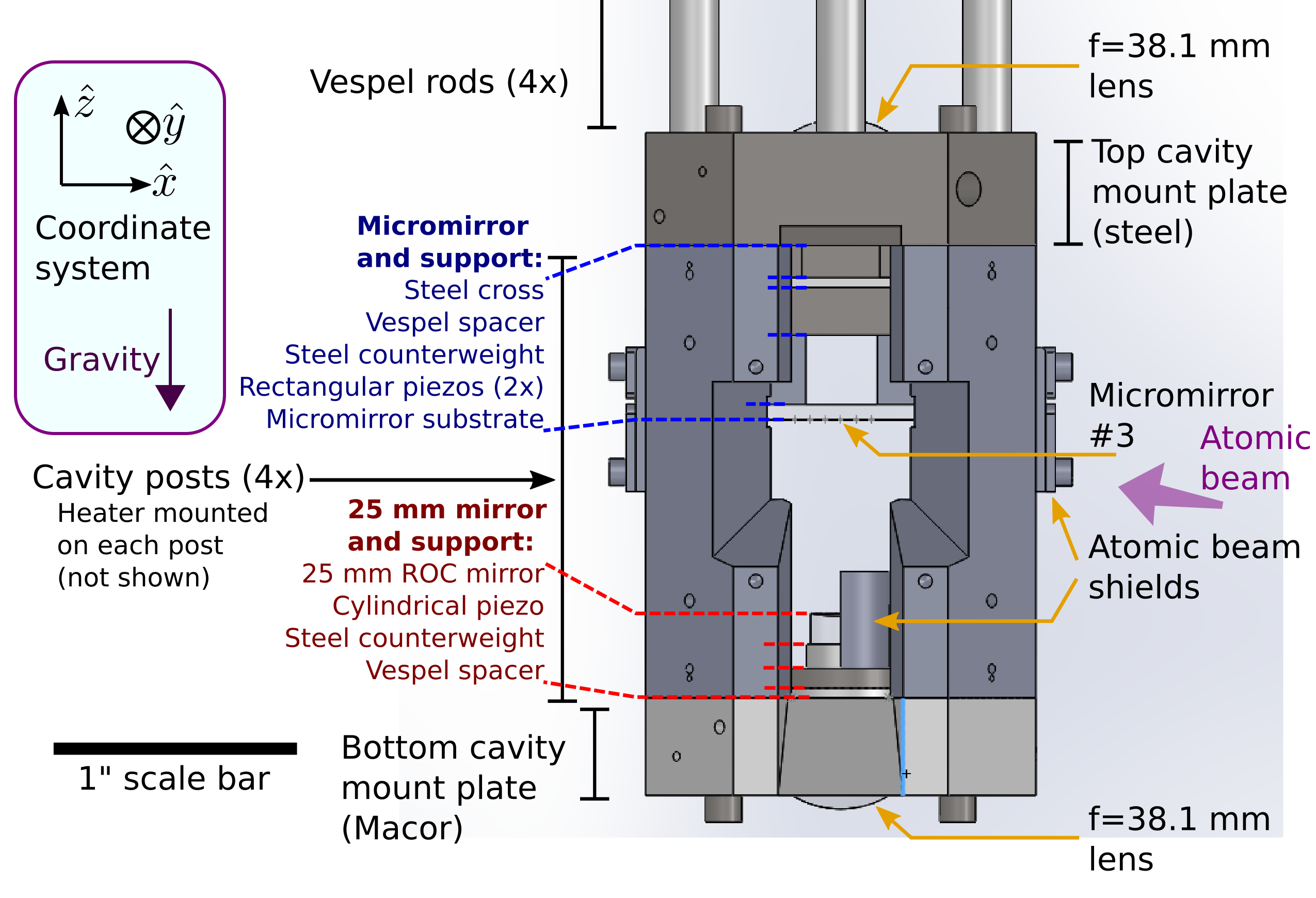}
 \caption{Structure of the asymmetric cavity: piezo is the short for piezoelectric actuator. Vespel is a polyimide-based plastic.} 
 \label{CavityPicture}
 \end{center}
\end{figure}

The mechanical structure supporting the cavity is shown in Fig. \ref{CavityPicture}. Its main part consists of mounting plates for the micromirror (top) and the 25-mm-ROC mirror (bottom), connected by four posts. The structure is made of type 316 stainless steel for mechanical strength and small magnetization, except for the bottom mount plate made of macor glass-ceramic to prevent eddy currents over the whole structure when the magnetic field is switched. The posts are designed to be as thick as possible to have stiff connections between top and bottom mount plates, with openings to ensure a large enough optical access to the atoms. The cavity mirrors are mounted on piezoelectric actuators (PZTs). The 25 mm ROC mirror has a 0.125 in. long single-layer PZT (Channel Industries material C5700) for fast tuning, and the micromirror substrate is attached to two 6.5 $\mu$m travel range, 9 mm long multistack PZTs (PI P-885.11) for slow but long-range tuning. Between each PZT and its mounting plate, a counterweight made of stainless steel and a damping layer made of polyimide-based plastic (VESPEL) is located to fully utilize the tuning of PZT for moving mirrors, without transmitting vibrations to the mounting structure. The top mount plate is suspended by thin VESPEL rods, in the middle of which stainless steel 4-40 screws tighten the cavity structure onto an adapter to a reducing flange, to dampen the vibrations from the environment through the adapter. 

The 556  and 759 nm light is sent to the cavity from the bottom side, after proper mode shaping. The cavity length is tuned over short distances by the PZTs, and over long distances by adjusting the temperature of the whole cavity mount, which is stabilized by a servo circuit. Each pillar has its own heater, and heaters can be controlled independently. This large tuning is important to have the cavity simultaneously resonant for 556 nm light on the $\mathrm{6s^2 \hspace{0.1 em} {}^{1} \hspace{-0.05 em} S_0 \rightarrow 6s6p \hspace{0.1 em}{} ^{3} \hspace{-0.05 em} P_1}$ transition and 759 nm light close to the magic wavelength for the 578 nm clock transition. In addition, the independent control of the four heaters enables the fine tuning of the tilt between two mirrors, which plays an essential role in maximizing the finesse of the cavity at a given $L$. 

The  cavity is locked to the 759 nm laser by Pound-Drever-Hall (PDH) locking \cite{ApplPhysB.31.97}. To perform the frequency stabilization of the cavity, feedback is applied to the short PZT, which has a bandwidth of 6 kHz, limited by a mechanical resonance of the cavity holding structure. To complement the small tuning range of the short PZT, the long PZT is tuned by another servo circuit with a $\sim 1$ Hz bandwidth to compensate for the long term cavity length drift, in excess of the length tuning possible by the short PZT. The cavity resonance frequency near 556 nm is tuned into resonance with the $\mathrm{6s^2 \hspace{0.1 em} {}^{1} \hspace{-0.05 em} S_0 \rightarrow 6s6p \hspace{0.1 em}{} ^{3} \hspace{-0.05 em} P_1}$ atomic transition by adjusting the 759 nm light frequency. To trap atoms for a long time \cite{PhysRevA.56.R1095}, the intra cavity 759 nm light is intensity stabilized with a bandwidth of $\sim 1$ MHz. This provides $\sim10$ dB suppression of the intensity noise of the intra-cavity light at $\sim 100$ kHz.

\begin{figure}[!t]
	\begin{center}
 \includegraphics[width=1\columnwidth0]{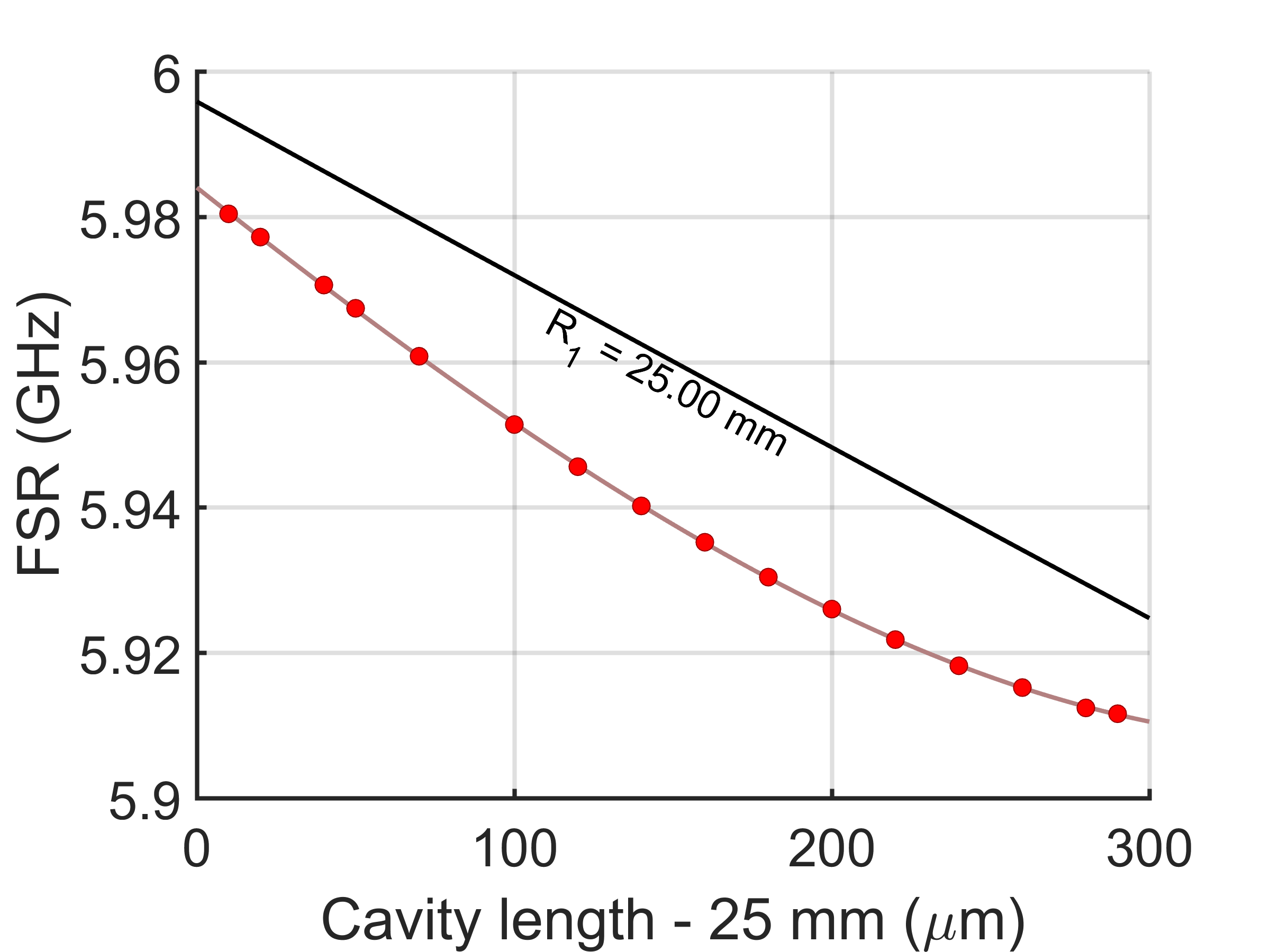}
 \caption{FSR for different mirror separation $L$: horizontal axis shows the mirror distance based on the calibration with flat-large ROC cavity, and vertical axis is corresponding FSR for each mirror distance. The thick black line shows the expected FSR using ${\rm FSR}=c/2L$ assuming $R_1$ is 25.00 mm. The red measured points (circles) can be fitted with a cubic function of ${\rm (FSR/[GHz])}=5.9840-0.345 (L'/{\rm [mm]}) +0.150 (L'/{\rm [mm]})^2+0.615 (L'/{\rm [mm]})^3$, where $L'=L-25$ mm (thin line).} 
 \label{FSRvsL}
 \end{center}
\end{figure}

%\section{Free spectral range of the asymmetric cavity}\label{AssymCavFSR}
As mentioned in Section \ref{CavProperty}, different measurement methods yield different distances between two mirrors of the asymmetric cavity. When the mirror distance of 25 mm (equal to $R_1$, independently measured to equal $25.00\pm0.01$ mm) is calibrated by the disappearance of the stable cavity mode of the flat-large ROC cavity described in Appendix \ref{AppA}, and then the micromirror is moved by a specific amount by a translational stage with a precision of 1 $\mu$m (Thorlabs MBT616D), the mirror distance is recorded as $L=25.0467(10)$ mm. On the other hand, the measurement of FSR of 5970.04(4) MHz suggests $L=25.10807(17)$ mm. Figure \ref{FSRvsL} shows this discrepancy in terms of the measured FSR as a function of the distance between the two mirrors. The graph shows nonlinearity in the relation between the FSR and the cavity length, which clearly shows the deviation from the standard formula of ${\rm FSR}=c/2L$. The tight waist exhibited by the asymmetric cavity leads to deviations from the paraxial approximation, which could cause corrections to the relationship between the cavity length and the FSR. However, we expect these corrections to be largest when the waist is small, i.e. $L\approx25.00$ and $L\approx25.30$ mm. The measurements (Fig. \ref{FSRvsL}) produce the opposite behavior, leaving this phenomenon currently unexplained. 

\bibliographystyle{apsrev4-1}
\bibliography{Instrumentation}

\end{document}